# Evaporation-induced colloidal diffusiophoresis in electrolyte solutions

Dang Q. Duong,[a] Jung Hyun Kim,[b] Hanna Cho,[b] Sangwoo Shin[a,]*

[a] Department of Mechanical and Aerospace Engineering, University at Buffalo, The State University of New York, Buffalo, NY 14260, USA

[b] Department of Mechanical and Aerospace Engineering, The Ohio State University, Columbus, OH 43210, USA

* Corresponding author: sangwoos@buffalo.edu

## Abstract

Non-volatile solutes accumulate near a receding drying front, generating concentration gradients that can drive directional colloidal transport by diffusiophoresis. This dynamic spatial arrangement of colloidal particles could affect the microstructure and macroscopic properties of the dried film. Here, we combine experiments and theory to investigate polystyrene particles suspended in electrolyte solutions drying within a capillary tube. By varying the electrolyte composition and ambient relative humidity, we show that evaporation-induced ion gradients drive particles toward the liquid-gas interface and enhance interfacial accumulation, whereas slower drying at higher humidity weakens the gradient and particle migration. We use the heat balance integral method to predict the transient solute field for a receding interface and couple it to particle transport to capture the observed particle trajectories and accumulation trends. These results identify background electrolytes and drying rate as coupled controls on particle distribution and provide a predictive framework to optimize formulation and processing parameters for manufacturing slurry-based materials, such as battery electrodes and turbine coatings.

**Keywords:** diffusiophoresis; evaporation; colloids; coating; integral balance method

## 1. Introduction

Drying of a colloidal film plays a central role in coating processes for a wide range of applications, including lithium-ion (Li-ion) and solid-state battery electrodes,[1,2] photonic crystals,[3,4] thermal barrier coatings,[5,6] and painting.[7,8] The functionalities of these films depend not only on their constituents, including colloidal particles and other non-volatile components, but also on the structures that emerge across length scales as those constituents rearrange during drying.[9,10] Competition between evaporation and diffusion can cause the non-volatile constituents to accumulate near the liquid-gas interface, generating concentration gradients within the liquid film. The resulting concentration gradients in solute species (e.g., ions, polymers) can drive directional colloidal transport through diffusiophoresis.[11] This transport process can lead to stratification or dense accumulation zones, ultimately impacting the microstructure and properties of the dried film.[12]

The direction of colloidal migration by diffusiophoresis is determined by the properties of both the particles and the chemical species present in the solution. For charged particles in electrolyte concentration gradients, the migration direction depends on the particle surface charge and the diffusivity contrast between ionic species, commonly represented by a dimensionless factor $\beta = \frac{D_+ - D_-}{z_+ D_+ - z_- D_-}$ for a binary electrolyte, with $D_\pm$ and $z_\pm$ denoting the diffusivity and valence of cation (+) and anion (−), respectively. For instance, negatively charged particles migrate up the gradients of ions of negative $\beta$ (e.g., LiCl, NaCl).[13–17] Conversely, gradients of non-adsorbing polymers (e.g., polyethylene glycol, dextran) drive particles down the polymer concentration gradient, i.e., toward regions of lower concentration, regardless of particle surface charge.[12,18–21] This motion arises from steric repulsion and the resulting osmotic imbalance near the particle surface.

Because drying colloidal suspensions are central to many coating applications, a substantial body of literature has reported stratification of binary colloidal mixtures, including studies of how polymers influence particle distribution in polymer-particle mixtures.[19,22–24] For instance, by considering polymer coils as ideal smaller particles in the mixture following the Asakura-Oosawa model,[25] Sear and Warren theoretically predicted the formation of a colloidal exclusion zone near the drying interface. This exclusion zone arises from diffusiophoretic repulsion of particles away from the drying interface caused by polymer concentration gradients.[12] This prediction is consistent with experimental observations by Baesch et al., who reported a particle-free layer above a particle accumulation zone in a drying particle-polymer mixture,[26] although the authors did not attribute this phenomenon to diffusiophoresis. In the same year, Sear further predicted analytically that such an exclusion zone only occurs within a range of initial polymer concentrations, depending on the Péclet number of the system.[24] More recently, Hooiveld et al. investigated the effect of background salt on colloidal diffusiophoresis of large particles driven by smaller colloids, leading to self-stratification. However, they did not examine how salt gradients affect colloidal transport.[27] In related theoretical work, Xu et al. modeled diffusiophoretic colloids in a unidirectionally drying particle-electrolyte suspension under the influence of gravity and showed that evaporation-generated ion gradients can alter particle redistribution.[28] Nonetheless, direct experimental measurements of colloidal transport in drying particle-electrolyte mixtures remain limited.

In this work, we experimentally observe transient particle motion near a receding drying front and quantify its dependence on electrolyte species, concentration, and ambient relative humidity. We then use the heat-balance integral approach to derive the evolving solute distribution associated with the transient interface velocity and couple it to models of particle trajectories and accumulation. Together, the experiments and theory examine how evaporation-induced ion gradients redistribute colloids during drying.

## 2. Materials and Methods

Lithium chloride (LiCl) and sodium chloride (NaCl) were purchased from Sigma-Aldrich, and polystyrene (PS) particles were purchased from Bangs Laboratories. We prepared PS dispersions in deionized (DI) water and in electrolyte solutions. Negatively charged PS particles with an average diameter of 1 μm were used in all experiments at an initial particle volume fraction of 0.01%. Electrolyte solutions contained 1 or 5 mM LiCl or 1 mM NaCl. The zeta potentials of PS particles suspended in each electrolyte solution were measured by electrophoretic light scattering (Litesizer 500, Anton Paar).

Particle motion during evaporation was recorded at 10 seconds per frame using an sCMOS camera (K5, Leica) through an inverted microscope (DMi8, Leica) using both bright-field and fluorescence imaging. A 5× objective was used to obtain a larger field of view and record the overall evolution of each dispersion for 1 h, whereas a 20× objective was used to track individual particles near the drying front for 25 seconds. Each experiment was repeated at least three times, or until a reliably uniform receding speed of the drying front was obtained for each relative humidity condition.

## 3. Results and Discussion

### 3.1 Experimental observation of drying-induced colloidal diffusiophoresis in electrolyte solutions

To establish controlled drying environments, we used a drying chamber consisting of a polydimethylsiloxane (PDMS) channel fabricated by standard soft lithography[29] in which a glass capillary tube (inner diameter = 50 μm, length ~ 1 cm) was secured (Figure 1a). Immediately after filling the capillary tube with a colloidal suspension, one end was sealed with a UV-curable epoxy (NOA 81, Norland Optical Adhesives). Air with relative humidity controlled at 0.5%, 40±10%, or 70±5% was then supplied into the channel at 8 psi gauge pressure ($\Delta P_a$) to regulate the liquid evaporation rate inside the tube. Then, the motion of colloidal particles around the receding liquid-gas interface was captured using a fluorescent microscope (Figure 1b).

Drying-induced motion of colloidal particles under various experimental conditions is shown in Figure 2. In DI water, particles undergo predominantly passive Brownian fluctuations with negligible net displacement until they are gathered at the receding liquid-gas interface, forming a particle band (Figures 2a-c; Movie S1). On the other hand, in 1 mM LiCl evaporation generates an electrolyte concentration gradient that drives directional particle migration toward the liquid-gas interface. The particle trajectories in the kymographs identify this directed motion, where the pathlines collectively march upwards in LiCl solution whereas the pathlines are more or less flat in DI water (Figures 2d-f; Movie S2).

We identify that this drying-induced directional migration is sensitive to the ambient relative humidity, suggesting the evaporation rate sets the driving force for the directed particle motion. As water evaporates, non-volatile ions accumulate near the interface, establishing the necessary chemical gradients that facilitate directional transport via diffusiophoresis.[12,30] At elevated humidity, the slower drying front allows ions to diffuse farther into the liquid, leading to an attenuation of the gradient and a subsequent decrease in the velocity of the particles (Figures 2d-f). Additional experiments with a different type of ions (NaCl) and concentrations also showed similar trends (Figure S3 in the SI).

### 3.2 Analytical model and numerical simulations

Motivated by the experimental observations, we construct an analytical model for the evaporation rates and the resulting solute and particle dynamics. Since the capillary tube was long compared to the field of view, we modeled the observed region of the capillary as a one-dimensional (1-D), semi-infinite domain (Figure 1c). During sample preparation, partial drying created a small air gap between the opening of the capillary and the liquid-gas interface, denoted $l_0$. Because both the initial solute and particle concentrations in this work are relatively dilute and the interface remained far from the capillary opening through the 1-hour observation period, we assume the interfacial solute concentration remained below the phase separation threshold.[31] We therefore treat drying as a single-phase liquid moving boundary problem over the experimental duration.[32,33] Doing so allows for determining the drying speed by equating the receding speed of the interface to the loss of water by diffusion of vapor along the gas column,[34] which results in the following expression:

$$\rho_w \frac{dl}{dt} = \frac{D_v M_v P_{vs}(1-\mathrm{RH})}{\bar{R}Tl(t)}, \quad (1)$$

where $l(t)$ is the position of the liquid-gas interface as a function of time, $\rho_w$ is the density of the liquid, $D_v$ is the vapor molecular diffusivity, $M_v$ is the vapor molecular weight, $\bar{R}$ is the universal gas constant, $T$ is the experimental temperature (24°C), $P_{vs}$ is the saturated vapor pressure at the experimental conditions (3 kPa),[35] and RH is the relative humidity of supplied air (0.5%, 40%, or 70%).[34] Because we maintained $\Delta P_a = 8$ psi of air gage pressure against the vapor diffusing direction, the effective vapor diffusivity is estimated as $D_v = \frac{P_{atm}}{P_{atm}+\Delta P_a} \cdot D_{v0} \left(= 1.6 \times 10^{-5} \frac{m^2}{s}\right)$ according to the Chapman-Enskog theory,[36] where $D_{v0} = 2.5 \times 10^{-5} \frac{m^2}{s}$ is the vapor diffusivity in air at atmospheric pressure $P_{atm}$.

The position of the liquid-gas interface $l(t)$ is obtained by integrating equation 1 with the initial condition $l(t=0)=l_0$, which reads

$$l(t)=\sqrt{2At+l_0^2}, \tag{2}$$

where $A=\frac{D_v M_v P_{vs}(1-RH)}{\rho_w \bar{R}T}$ and $l_0 \approx 470$ μm based on the experimental trials. This expression for $l(t)$ is used to reconstruct the solute concentration profile, which ultimately governs the motion of the dispersed particles.

Since the drying velocity, $v_d(t)=\frac{dl}{dt}=\frac{A}{\sqrt{2At+l_0^2}}$, does not exactly scale with $t^{-\frac{1}{2}}$, a standard self-similar solution is not available for the solute distribution. We therefore use the mass balance integral method to approximate this moving boundary problem.[37] This approach is the mass transfer analog of the classical heat balance integral originally proposed by Goodman,[38] with solution diffusion substituting heat conduction.

We consider a 1-D, semi-infinite domain extending in the positive $x$ direction (Figure 1c), with $D_s$ being the solute diffusivity and $c(x,t)$ the solute concentration. The solute field is governed by Fick's second law,[39]

$$\frac{\partial c}{\partial t}=D_s\frac{\partial^2 c}{\partial x^2},\;\; x\in[l(t),\infty]. \tag{3}$$

The initial and boundary conditions are

$$c(x,0)=c_0,\;\; c(\infty,t)=c_0,\;\; \text{and}\;\; v_d c_i=-D_s\frac{\partial c}{\partial x}\Big|_{x=l(t)}, \tag{4}$$

where $c_0$ is the initial solute concentration and $c_i$ is the solute concentration at the liquid-gas interface. The coordinate system is then transformed to $\hat{x}=x-l(t)$, which yields an advection term in equation 3, such that

$$\frac{\partial c}{\partial t}-v_d\frac{\partial c}{\partial \hat{x}}=D_s\frac{\partial^2 c}{\partial \hat{x}^2}. \tag{5}$$

The corresponding boundary conditions at the liquid-gas interface are

$$c(\hat{x}=0,t)=c_i \;\text{and}\; v_d c_i=-D_s\frac{\partial c}{\partial \hat{x}}\Big|_{\hat{x}=0}. \tag{6}$$

Once the governing equation and boundary conditions are defined, the balance integral method begins with an introduction of a heat penetration depth,[38,40] which is otherwise denoted here as the solute penetration depth $\delta_c$. Diffusion is significant within $\delta_c$ and negligible beyond it. The mass balance integral equation is then constructed by multiplying both sides of equation 5 by $d\hat{x}$ and integrating from 0 to $\delta_c$, i.e., from $l(t)$ to $l(t)+\delta_c$ in the original $x$ coordinate. Following the framework established by Goodman[38] while assuming a cubic polynomial profile,[40] the solute concentration profile, $c(\hat{x},t)$, and the corresponding solute concentration at the liquid-gas interface, $c_i$, can be written as

$$c(\hat{x},t) = c_0 + (c_i - c_0)\left(1 - \frac{\hat{x}}{\delta_c}\right)^3, \tag{7}$$

$$c_i = \frac{\alpha + \sqrt{\alpha^2 - 4c_0^2}}{2}, \tag{8}$$

where $\alpha = c_0\left\{2 + \frac{4}{3D_s}v_d\left[l(t) - l_0\right]\right\}$ and $\delta_c = \frac{3D_s(c_i - c_0)}{c_i v_d}$. The analytical solute distribution predicted by the mass balance integral method (equation 7) is validated against numerical solutions obtained with MATLAB's built-in partial differential equation solver, pdepe.[41]

The analytical and numerical solutions for the solute concentration profiles are compared in Figures 3a-c using dimensionless variables $C = \frac{c(\hat{x},t)}{c_0}$, $X = \frac{x - l_0}{d} = \frac{\hat{x} + l(t) - l_0}{d}$, and $\tau = \frac{t}{t_{ref}}$, where $t_{ref} = \frac{\rho_w \bar{R} T d^2}{D_v M_v P_{vs}}$ is the characteristic evaporation time scale obtained from equation 1 and $d$ is the inner diameter of the capillary. Overall, the close agreement between the two validates the cubic approximation for the solute field. As water evaporates at the liquid-gas interface, non-volatile solute accumulates near the interface and generates solute gradients along the liquid column (Figure 1c). The magnitude of this buildup depends on the drying velocity, which is inversely proportional to the ambient relative humidity, as described in equation 1. In Figures 3a-c, the interfacial concentration $c_i$ increases continuously and drives diffusiophoretic transport. Particularly, at a low relative humidity of RH = 0.5%, salt concentration at the interface, $c_i$, reaches 1.7 times the initial concentration $c_0$ after $512\tau$ ($\approx$ 1 hour), whereas at RH = 70%, the decelerated recession of the interface allows ions to diffuse farther back into the capillary, limiting the accumulation to only $1.25c_0$ after $512\tau$. This trend is consistent with the weaker particle migration observed experimentally at elevated humidity.

To quantify the evaporation-induced directional particle transport, we evaluate the diffusiophoretic velocity $\boldsymbol{u_{dp}}$ and mobility $M_{dp}$. An electrolyte concentration gradient drives the particle at velocity[20]

$$\boldsymbol{u_{dp}} = M_{dp} \nabla \ln c. \tag{9}$$

Using the solute concentration profile in equation 7, the expression for $\boldsymbol{u_{dp}}$ in the 1-D problem reads

$$u_{dp} = \frac{-3(c_i - c_0)\left(1 - \frac{\hat{x}}{\delta_c}\right)^2 M_{dp}}{\delta_c\left[c_0 + (c_i - c_0)\left(1 - \frac{\hat{x}}{\delta_c}\right)^3\right]}. \tag{10}$$

For a binary symmetric electrolyte ($|z_+| = |z_-| = z$) in the thin Debye layer limit, the particle diffusiophoretic mobility may be expressed as

$$M_{dp} = \frac{\epsilon k_B T}{\eta e}\left\{\beta\zeta_p - \frac{2k_B T}{z^2 e}\ln\left[1 - \tanh^2\left(\frac{ze\zeta_p}{4k_B T}\right)\right]\right\}, \tag{11}$$

where $\epsilon$ is the permittivity of the medium, $\eta$ is the viscosity of the medium, $k_B$ is the Boltzmann constant, $T$ is the temperature, $e$ is the elementary charge, and $\zeta_p$ is the zeta potential of the particle.[13] While this expression works generally well for ideal, weak gradients, which has been used in numerous experimental studies involving colloidal diffusiophoresis driven by electrolyte gradients,[15,42–47] it often fails to capture non-ideal effects such as varying zeta potential and finite Debye layer effects.[11] Therefore, instead of using the above mobility expression, we determine $M_{dp}$ directly by fitting the predicted velocity from equation 10 to particle velocities extracted from individual particle tracking using Trackmate,[48] which will be discussed in Section 3.3.

Particle trajectories $x_p(t)$ were then obtained by numerically integrating $\frac{dx_p}{dt} = u_{dp}$. As illustrated in Figures 3d-f, 12 particles initialized at evenly spaced positions along the $x$ direction move toward the receding liquid-gas interface under the influence of the LiCl concentration gradient. At higher relative humidity, the predicted trajectories flatten as the driving force weakens. The model quantitatively reproduces the experimental pathlines without free parameters, supporting its use for describing particle transport in the drying dispersion.

The diffusiophoretic velocity estimated from equation 10 is also used to predict the particle number density field $n(x, t)$ by solving the advection-diffusion equation numerically[13]

$$\frac{\partial n}{\partial t} = D_p \frac{\partial^2 n}{\partial x^2} - \frac{\partial}{\partial x}(u_{dp} n), \tag{12}$$

where $D_p$ is the particle diffusivity given by the Stokes-Einstein equation, $D_p = \frac{k_B T}{6\pi\mu r_p}$, with $r_p$ is the particle radius. At the onset of evaporation ($t = 0$), $N = \frac{n}{n_0} = 1$, where $n_0$ is the initial particle number density. A no-flux condition at the interface $\left(\frac{dn}{d\hat{x}}\Big|_{\hat{x}=0} = 0\right)$ is implemented to prevent particles from escaping the liquid column. Full numerical details, verification, and convergence are provided in the SI.

The normalized particle profiles in Figures 4a-c show rapid particle accumulation toward the interface driven by electrolyte gradients, with weaker accumulation at higher relative humidity, which is consistent with the experimental observations. To assess the effects of different solute species and concentration on the particle transport, we introduce the diffusiophoretic Péclet number $\left(Pe_{dp} = \frac{M_{dp}}{D_p}\right)$ and the solute Péclet number $\left(Pe_s = \frac{\bar{v}_d h_c}{D_s}\right)$, where $\bar{v}_d$ is the average velocity of the moving front over $50\tau$, $h_c$ is the length of the physical domain, and $D_s$ is the electrolyte ambipolar diffusivity, given by $D_s = \frac{(z_+ - z_-) D_+ D_-}{z_+ D_+ - z_- D_-}$.[13]

Because the same particles are used in all experiments, $Pe_s$ only varies with $\bar{v}_d$, which depends on RH. Across RH values from 0.5% to 70%, the receding interface sweeps particles toward the boundary, producing moderate to strong accumulation. As depicted in Figures 4d-f, particle accumulation at the interface is greatest when a high $Pe_s$, associated with slow solute diffusion, is combined with high $Pe_{dp}$, associated with strong diffusiophoretic mobility. The predicted interfacial particle accumulation $N_i$ across a wide range of RH (Figure 4g) shows that the background electrolyte enhances particle enrichment at the liquid-gas interface.

To gauge the contribution of diffusiophoresis to particle accumulation, we define $\Gamma = \frac{N_{i,\text{salt}}}{N_{i,\text{DIW}}}$ as the ratio of interfacial particle accumulation with electrolyte to that in DI water, and $\Delta = \frac{w_{\text{salt}}}{w_{\text{DIW}}}$ as the ratio of the width of the particle accumulation layer ($w$) with electrolyte t to that in DI water, thereby isolating the accumulation associated with diffusiophoresis from the accumulation by the direct sweeping of the receding drying front. As shown in Figure 4h, 1 mM LiCl solution recruits significantly more particles to the interface than 1 mM NaCl or 5 mM LiCl over the same duration. Moreover, the existence of electrolyte gradients in the drying solution creates an accumulation layer that is 50% more compressed than that being swept alone by the drying front (Figure 4i). Particularly, LiCl compresses such a layer roughly 20% more than NaCl does. The calculation process of this accumulation layer is presented in the SI. This trend reflects the concentration-dependent particle zeta potential and the larger diffusivity contrast magnitude $|\beta|$ of LiCl. For all electrolyte conditions, the relative diffusiophoretic enhancement peaks at approximately 80-90% RH and then decreases sharply as the drying rate becomes too low to sustain substantial solute gradients.

### 3.3 Two-dimensional assessment of particle velocity and mobility in electrolyte solution

Drying-induced electrolyte gradients can not only drive particle diffusiophoresis, but also generate fluid flow by diffusioosmosis, a complementary transport mechanism occurring at the capillary wall. In a confined dead-end pore, diffusioosmotic wall slip induces a pressure-driven backflow and a recirculating axial velocity profile that varies with radial position (Figure 5a).[13] The recirculating flow field can be found by solving the Stokes equations with a diffusioosmotic slip velocity at the wall $u_{f,x}\big|_{r=\frac{d}{2}} = M_{do}\frac{\partial \ln c}{\partial x}$, where $M_{do}$ is the wall diffusioosmotic mobility. The axial fluid velocity $u_f$ in the $x$ direction is[13]

$$u_f(r,x) = M_{do}\frac{\partial \ln c}{\partial x}\left(\frac{8r^2}{d^2} - 1\right), \tag{13}$$

so that the particle velocity $u$ is a result of the sum of diffusiophoretic velocity $u_{dp}$ and fluid velocity $u_f$, i.e., $u = u_{dp} + u_f$.

While the particle velocity is a sum of diffusiophoresis and circulating background flow, the flow vanishes at $r = \pm\frac{d}{\sqrt{8}}$, thus measurements at these radial positions isolate the diffusiophoretic contribution (Figure 5b). The remaining radial variation is then used to determine the wall diffusioosmotic mobility (Figure 5c-e). We note that the diffusioosmotic contribution to the particle trajectories (Figure 3d-f) and concentration distributions (Figure 4a-c) is neglected in the 1-D model because of the zero net volumetric flow.

The particle diffusiophoretic mobilities $M_{dp}$ in 1 mM LiCl and 1 mM NaCl solutions obtained from tracking individual particles (Figure 5b) as well as the corresponding wall diffusioosmotic mobilities $M_{do}$ obtained by fitting equation 13 are presented in Table 1. At the same concentration, the particle mobility is higher in LiCl than in NaCl, consistent with the fact that the diffusivity contrast $|\beta|$ for LiCl ($\beta_{\mathrm{LiCl}} = -0.33$) is greater than that for NaCl ($\beta_{\mathrm{NaCl}} = -0.21$). This difference promotes faster accumulation at the interface, consistent with the predictions in Figure 4. Using the measured zeta potentials $\zeta_p$ in the theoretical mobility expression from equation 11 yields diffusiophoretic mobilities of 725 $\mu m^2/s$ and 583 $\mu m^2/s$ for particles dispersed in 1 mM LiCl and 1 mM NaCl, respectively, which are about 5% higher than the corresponding fitted values in Table 1. This discrepancy may arise because the zeta potential measured at a uniform bulk concentration does not truly represent the local value experienced by particles in the evolving concentration field, where the increased local ionic strength near the drying front can reduce the magnitude of the zeta potential and therefore the mobility.

## 4. Conclusion

Evaporation from a colloidal dispersion concentrates non-volatile solute near the receding interface, generating a concentration gradient that drives directional particle transport by diffusiophoresis. We have shown experimentally and theoretically that particles migrate toward the drying front and that the migration depends on solute identity, solute concentration, and ambient relative humidity. A larger magnitude of $\beta$ in LiCl produced stronger interfacial accumulation than the smaller magnitude of $\beta$ in NaCl. As relative humidity increased, the interface velocity and solute Péclet number decreased, which led to accumulated ions diffusing farther into the liquid and weakening the concentration gradient, thereby decreasing the particle migration speed and the interfacial particle accumulation. The coupled transport model successfully reproduced the observed trends in particle trajectories and accumulation. These results establish a mechanistic link between drying conditions, electrolyte properties, and particle redistribution near a receding interface, which guides future studies that connect particle transport to the microstructure and performance of dried slurry coatings. A relevant application is Li-ion battery anode manufacturing, where styrene–butadiene rubber (SBR) and sodium carboxymethyl cellulose (CMC) are widely used together as an aqueous binder system.[49,50] Our findings suggest that controlling ambient humidity and dissolved electrolyte composition potentially improves the microstructural homogeneity in Li-ion battery electrodes.

## Acknowledgments

This work was funded by the National Science Foundation via grant no. 2237177 and the ACS Petroleum Research Fund under New Directions Grant 69700-ND9.

**Table 1.** Experimental diffusiophoretic and diffusioosmotic mobilities of particles and fluid flow, respectively, in LiCl and NaCl solutions.

| Salt | LiCl | NaCl |
| --- | --- | --- |
| $M_{dp}\ \left(\frac{\mu m^2}{s}\right)$ | $685.71 \pm 20.06$ | $554.44 \pm 12.57$ |
| $M_{do}\ \left(\frac{\mu m^2}{s}\right)$ | $1266.67 \pm 149.63$ | $1207.56 \pm 98.99$ |

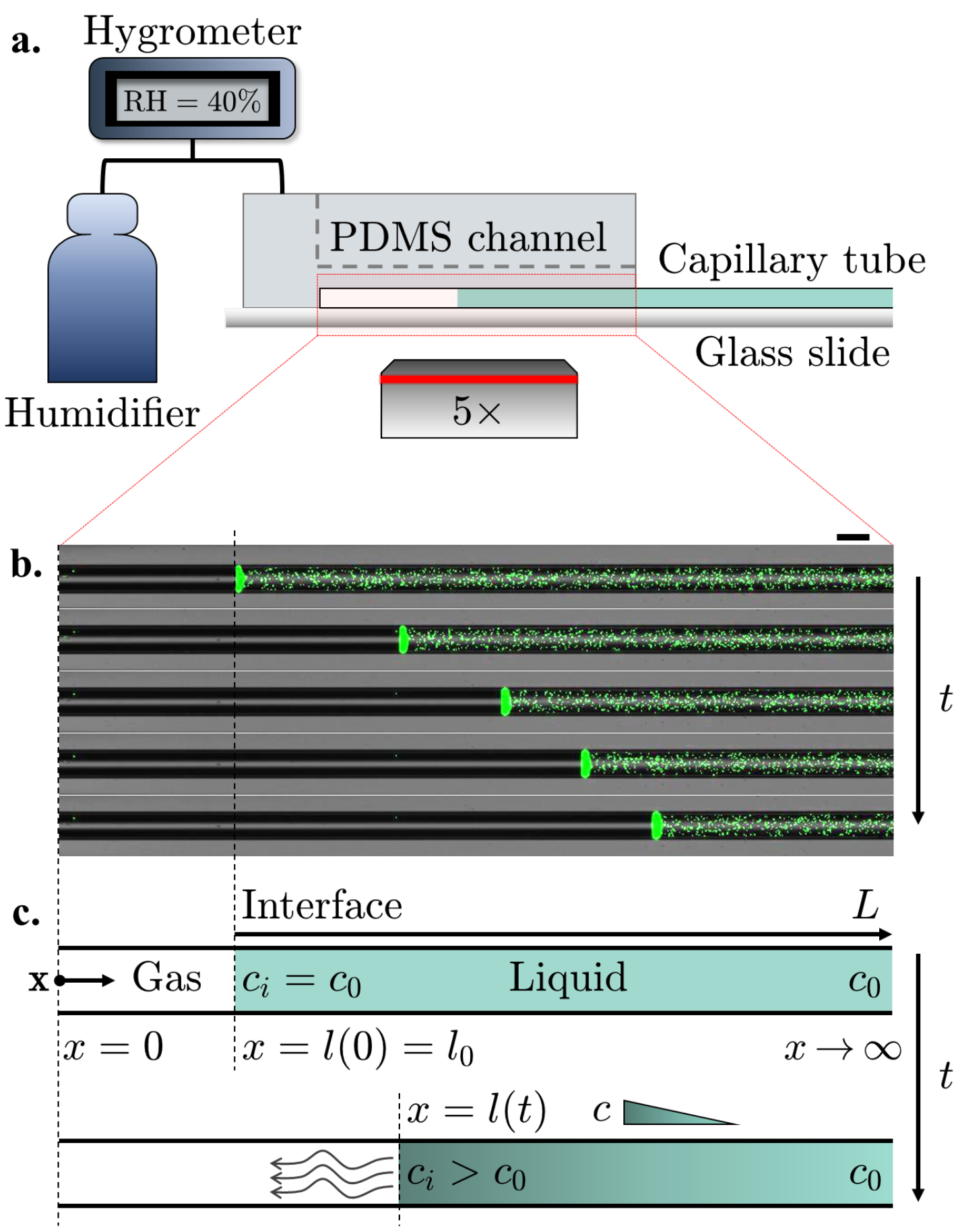

**Figure 1. Experimental setup and model domain.** (a) A circular capillary tube (~1 cm long; 50 µm inner diameter; 80 µm outer diameter) was secured inside a channel made of PDMS. Immediately after filling colloidal dispersion, the right end was sealed with a UV-curable epoxy, and the left end was exposed to a controlled airflow. The supplied air was humidified to control the evaporation rate within the capillary tube. The air pressure was kept constant at 8 psi. (b) Evaporation concentrates non-volatile solute (1 mM LiCl in this figure) near the interface, generating a concentration gradient that drives particles toward it. (c) The capillary is modeled as a 1-D, semi-infinite domain. $c_0$ denotes the initial solute concentration. The initial interface position, $l_0$, is established during sample preparation; the interface then recedes in the positive $x$ direction. The initial particle volume fraction is $\phi_0 = 0.01\%$. The scale bar is 100 µm.

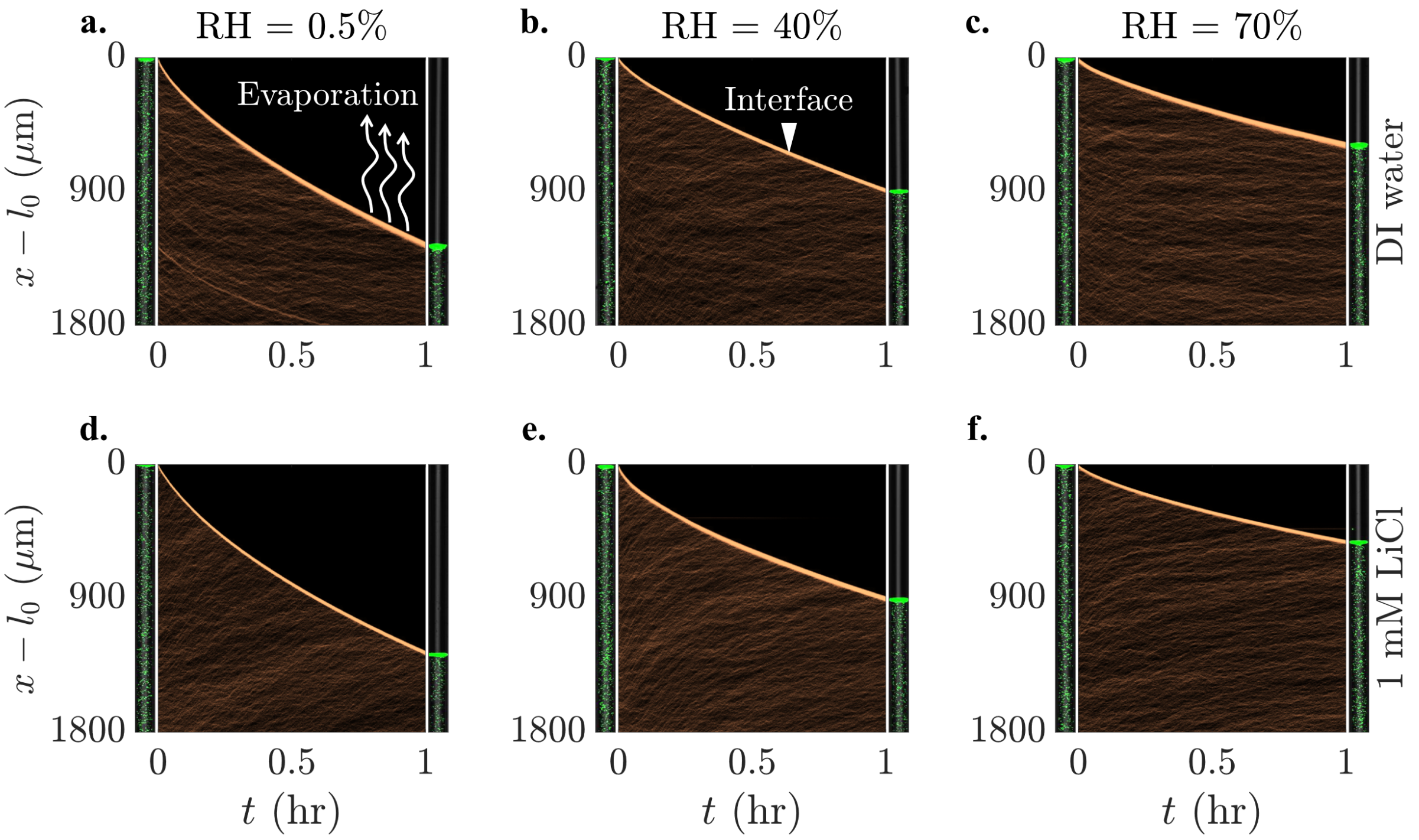

**Figure 2. Particle dynamics during drying at different relative humidities.** Kymographs generated from particle movies along with snapshots of the capillary at $t = 0$ and 1 hr. (a-c) In DI water, particles undergo predominantly Brownian motion. (d-f) In 1 mM LiCl, evaporation generates an ion concentration gradient that drives negatively charged PS particles toward the liquid-gas interface by diffusiophoresis. As relative humidity (RH) increases from 0.5% to 70%, the interface recedes more slowly, resulting in weaker gradients that produce flatter particle trajectories.

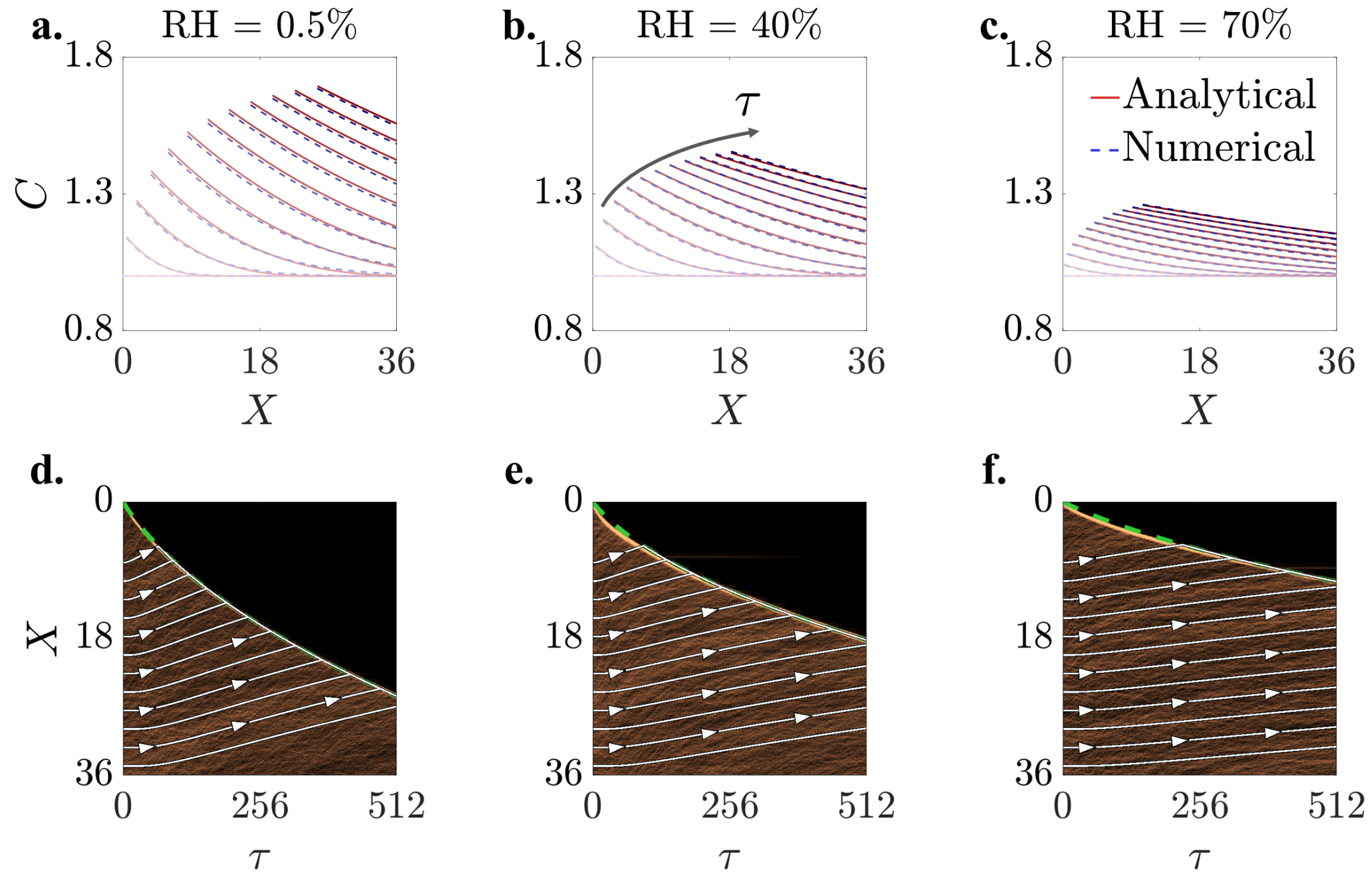

**Figure 3. Analytical and numerical predictions of solute and particle dynamics.** Columns correspond to RH of 0.5% (a,d), 40% (b,e), and 70% (c,f). (a-c) Analytical (solid red; equation 7) and numerical (dashed blue) LiCl concentration profiles are shown at successive times. Higher RH allows ions to diffuse farther into the liquid and weakens the concentration gradient. (d-f) Experimental kymographs and predicted particle pathlines show negatively charged particles moving toward the receding interface under the LiCl gradient. The green dashed line marks the interface position from equation 2.

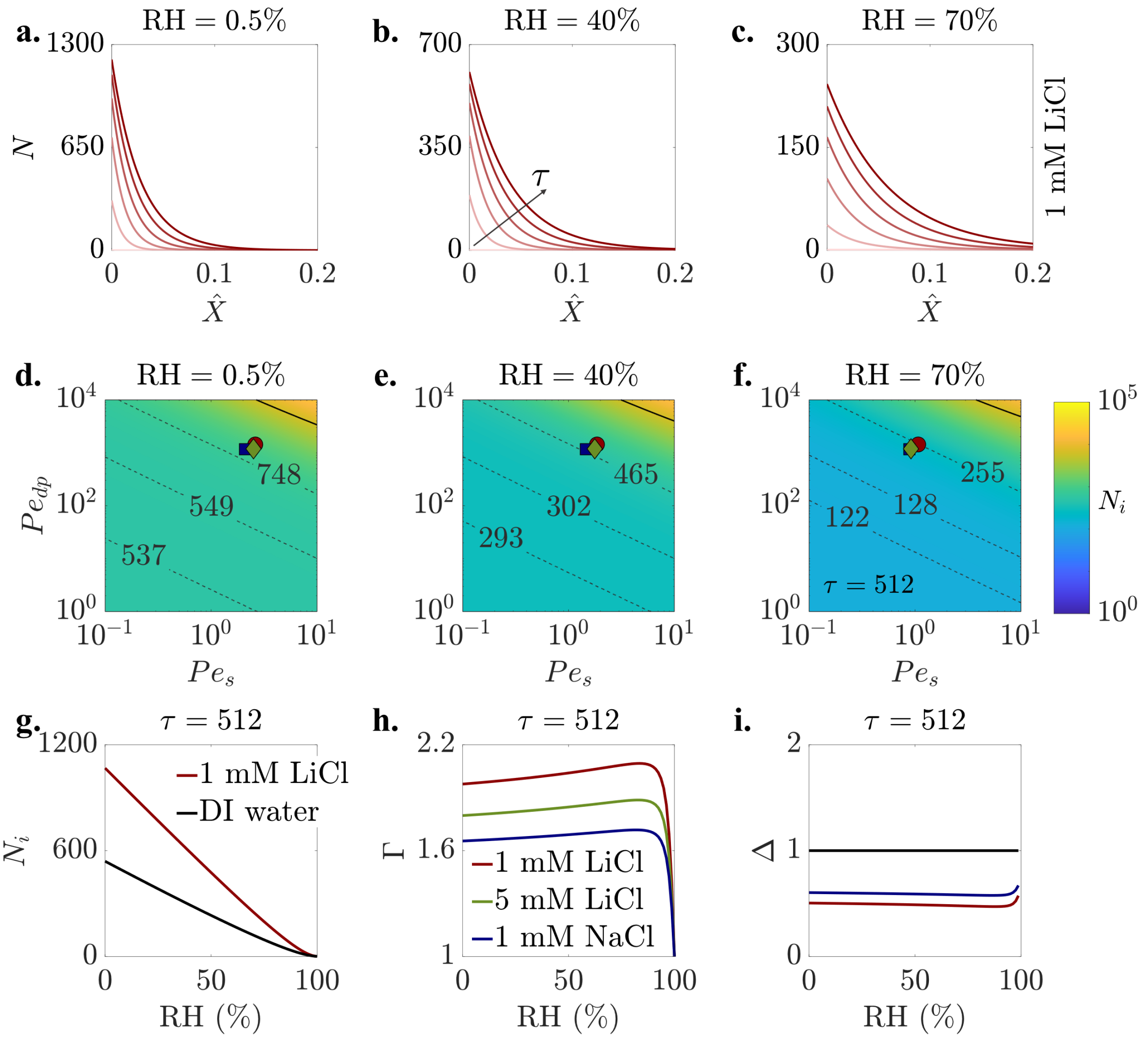

**Figure 4. Predicted particle accumulation driven by electrolyte gradients.** (a-c) Particle density profiles at varying RH. Increasing RH from 0.5% to 70% reduces interfacial accumulation by approximately 75%. (d-f) Interfacial accumulation $N_i$ after $512\tau$ versus the solute Péclet number $Pe_s$ and the diffusiophoretic Péclet number $Pe_{dp}$ at each RH. Larger $Pe_s$ and $Pe_{dp}$ jointly increase interfacial accumulation. Symbols indicate experimental conditions (red circles, 1 mM LiCl; green diamonds, 5 mM LiCl; blue squares, 1 mM NaCl). The black solid line indicates the packing limit of 64% beyond which steric interactions should be taken into account when estimating the particle density field at the interface. (g) 1 mM LiCl increases interfacial accumulation by nearly 100% relative to DI water. (h,i) Relative diffusiophoretic enhancement versus RH for various electrolytes. (h) $\Gamma$ compares the interfacial accumulation in electrolyte to that in DI water. (i) $\Delta$ represents the ratio of the width of the particle accumulation layer in electrolyte to that in DI water. Electrolyte gradients compress the particle accumulation layer almost twice as much as does DI water.

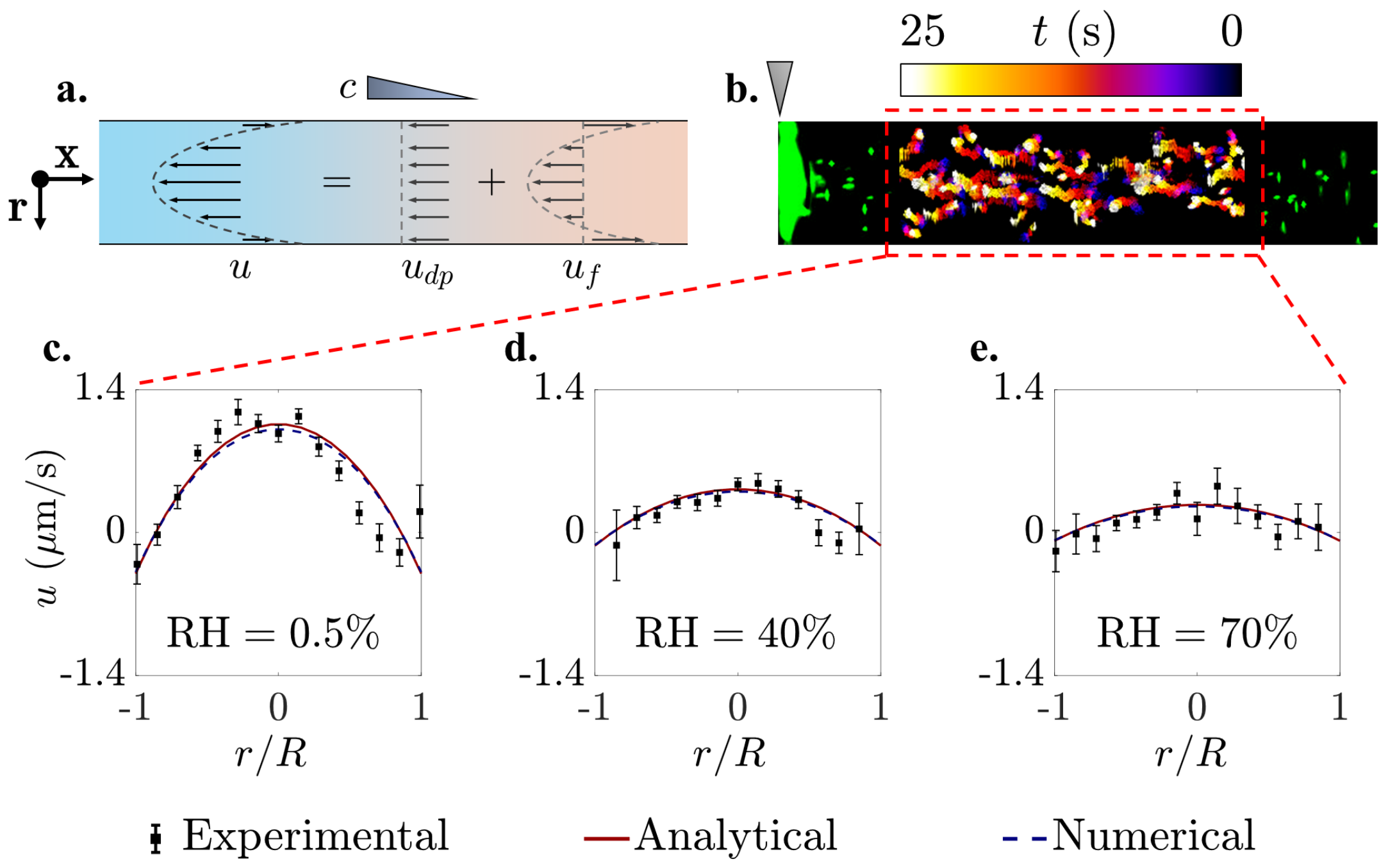

**Figure 5. Two-dimensional particle motion in 1 mM LiCl influenced by diffusiophoresis and diffusioosmosis.** (a) Particle velocity field ($u$) induced by diffusiophoresis ($u_{dp}$) and a recirculating flow ($u_f$) driven by diffusioosmotic slip and resulting back-pressure. (b) Particle trajectories recorded over 25 s within a region of interest approximately 65 µm from the interface indicated by the red dashed box. The color code represents time. The inverted gray triangle indicates the position of the liquid-gas interface. The scale bar represents 25 µm. (c-e) Axial velocity profiles at varying RH values. Points and error bars show binned measurements and standard deviations; curves show the fits from equation 13. The velocity at $r = \pm d/\sqrt{8}$ yields the particle diffusiophoretic mobility, thus the remaining radial variation yields the wall diffusioosmotic mobility (Table 1)